\documentclass[rsc,pccp,showpacs,superscriptaddress,amsmath,twocolumn]{revtex4}

\usepackage{graphicx}
\usepackage{bm} 
\usepackage{gensymb} 
\usepackage[percent]{overpic} 

\begin{document}

\title{Structural transformation between long and short-chain form of liquid sulfur \\ from \textit{ab initio} molecular dynamics}

\author{Du\v{s}an Pla\v{s}ienka}

\email{plasienka@fmph.uniba.sk}

\affiliation{Department of Experimental Physics, Comenius University, Mlynsk\'{a} Dolina F2, 842 48 Bratislava, Slovakia}

\author{Peter Cifra}

\affiliation{Polymer Institute, Slovak Academy of Sciences, D\'{u}bravsk\'{a} cesta 9, 845 41 Bratislava, Slovakia}

\author{Roman Marto\v{n}\'{a}k}

\affiliation{Department of Experimental Physics, Comenius University, Mlynsk\'{a} Dolina F2, 842 48 Bratislava, Slovakia}

\pacs{61.20.-p, 61.25.-f, 62.50.-p, 64.70.Ja}




 

\date{\today}

\begin{abstract}

We present results of \textit{ab initio} molecular dynamics study of a structural transformation occurring in hot liquid sulfur under high pressure, which corresponds to the chain-breakage phenomenon recently observed experimentally by Liu \textit{et al.} \cite{Liu-S} and to the electronic transition reported by Brazhkin \textit{et al.} \cite{Brazhkin-S, Brazhkin-S-Se-I}. We performed an extensive \textit{ab initio} study and confirmed the existence of one transformation separating two distinct liquid polymeric phases: one composed of short chain-like fragments and another one with very long chains. We have not observed additional transformations reported in Refs. \cite{Brazhkin-S, Brazhkin-S-Se-I} and in the recent theoretical study by Zhao and Mu \cite{Zhao-Mu} and our findings are in agreement with the most recent experiment \cite{Liu-S}. We offer a structural description of this liquid-liquid transformation in terms of chain lengths, cross-linking and geometry and investigate several physical properties. We conclude that the transformation is accompanied by changes in configurational energy and entropy as well as in diffusion coefficient and electronic properties (semiconductor-metal transition), but no density discontinuity was found. We also describe analogy of the investigated phenomenon to similar processes present in two other chalcogenes selenium and tellurium. Finally, we discuss possibility of a critical point terminating the studied transition in relation to the long-known behavior of ambient-pressure heated sulfur melt.
\end{abstract}

\maketitle

\section{Introduction}

Liquid-liquid transition (LLT) of a first-order kind in a stable liquid state was first unambiguously observed in phosphorus \cite{Katayama-P}
and since then it has been systematically searched for in various stable and undercooled liquids \cite{Katayama-Tsuji, McMillan-AAT-review-1, WWMcMillan-AAT-review, Machon-PIA-AAT-LLT-review, Franzese-LLT}.
LLT is an interesting thermodynamic phenomenon, which might exhibit parallelism with the structural evolution in solid state (crystal-crystal or amorphous-amorphous transition) as far as short-range order of liquids and solids can in certain cases be associated.
Transformations in liquids may be induced by changing pressure (transition between high-density liquid and low-density liquids) or by changing temperature in case of transitions between high-temperature liquid (HTL) and low-temperature liquid (LTL).
LLT is more or less sharply exhibited in stable state of several simplest systems like P \cite{Katayama-P, Hohl-Jones-P, Monaco-P, Ballone-Jones-P}, Ce \cite{Cadien-Ce},
S \cite{Brazhkin-S, Brazhkin-S-Se-I, Liu-S, Zhao-Mu}, Se
\cite{Warren-Dupree, Tamura, Kirchhoff-1, Kirchhoff-2, Kirchhoff-3, Kresse-Se, Shimojo-Hoshino-Zempo, Shimojo-Se, Hoshino-Shimojo, Shimojo-Se-2, Brazhkin-S-Se-I, Katayama-Tsuji, Katayama-Se, Brazhkin-Se},
Te \cite{Tsuchiya, Akola-Jones-1, Akola-Jones-2} and N \cite{Radousky, Mazevet, Ross-Rogers, Mukherjee-Boehler, Goncharov-N-2008, Boates-Bonev-N-2009, Donadio, Boates-Bonev-N-2011}
and is also predicted in H \cite{Scandolo-H-LLT}, C \cite{Glosli-Ree} or CO$_2$ \cite{Boates-CO2-LLT-1, Boates-CO2-LLT-2} as well as in numerous liquid metals
\cite{Franzese-LLT} and other composite systems. In some other materials, rather sharp changes were predicted to occur in their supercooled regimes, including water \cite{Poole-H2O} and silicon \cite{Sastry-Angell-Si}. Additionally, many liquids undergo gradual changes of their structure, which are usually not considered as genuine phase transitions.
An example of this case is the well-known $\lambda$-transition in sulfur.

Sulfur's structure belongs to the most complicated of all elements with numerous cyclic and chain-like molecular conformations
\cite{Donohue, Meyer, Steudel-S-review, Wong-S-review, Eckert-Steudel-S-review, Greenwood-S-review, Wong-Steudel-Steudel, Cioslowski, Hohl, Jones-Ballone-1}
and a large number of solid allotropes, some with very complex structure \cite{Meyer, Steudel-S-review, Wong-S-review, Eckert-Steudel-S-review, Greenwood-S-review, Donohue}.
The ambient-conditions stable form of sulfur is insulating $Fddd$ phase S-I ($\alpha$-S) with 16 ring-shaped S$_8$ molecules (128 atoms) arranged in
orthorhombic unit cell. Other stable molecular phases include orientationally disordered monoclinic $\beta$-S and $\gamma$-S made of S$_8$ units and phase S-VI
composed of molecules S$_6$. On rising pressure, S-I polymerizes into semiconducting chain phases - trigonal S-II \cite{Crichton, Deg-1, Deg-2, Deg-5} and tetragonal
S-III \cite{Fujihisa, Deg-1, Deg-2, Deg-5}, which are composed of infinite parallel helical chains with different geometry. Further compression leads to extended-network
phases - incommensurately modulated $bcm$ phase S-IV \cite{Hejny, Deg-3, Deg-4} and S-V with $\beta$-Po structure \cite{Luo-Greene-Ruoff, Deg-3, Deg-4, Deg-5}, which are both
metallic \cite{Deg-4, Nishikawa, NNSO} and superconducting \cite{Gregoryanz-supercond}.
Other structures of S were predicted from computer simulations - $\alpha^{\prime}$-S \cite{Pastorino-Gamba-2}, monoclinic-S with two different conformations
of S$_8$ molecules \cite{Plasienka-Martonak-S-PIA}, chain forms similar to S-II and S-III \cite{Oganov-Glass-USPEX} or higher-pressure superconducting $sc$ and $bcc$
structures \cite{Zakharov-Cohen, Rudin-Liu}. Besides the crystalline regime, sulfur also undergoes pressure-induced amorphization \cite{Luo-Ruoff, Luo-Greene-Ruoff, Akahama, Sanloup, Plasienka-Martonak-S-PIA}
and the possibility of amorphous-amorphous transition (polymorphism in glassy state) \cite{Sanloup} has also been proposed.

Sulfur at ambient pressure melts around 115$\degree$C \cite{Meyer} and then undergoes a transformation from predominantly molecular liquid into a
polymeric state at around $T_\lambda$=159$\degree$C, which is accompanied by sharp changes in density, heat capacity and viscosity \cite{Meyer, Steudel-Eckert-S-review, Eckert-Steudel-S-review}.
This is the well-known $\lambda$-transition, in which changes in sulfur structure proceed continuously with a wide coexistence region of molecular and polymeric units.
It is therefore usually understood as a kinetically-controlled equilibrium reaction between small cyclic and long polymeric molecules in a model developed by
Tobolsky and Eisenberg \cite{Steudel-Eckert-S-review, Eckert-Steudel-S-review, Tobolsky-Eisenberg}, or as a second-order phase transition \cite{Wheeler-Kennedy-Pfeuty}.
The process of temperature or photo-induced ring-opening polymerization in S$_8$ rings was studied theoretically by classical and \textit{ab initio} molecular dynamics
(MD) \cite{Stillinger-Weber-LaViolette, Tse-Klug-S, Shimojo-S, Munejeri} and Monte Carlo \cite{Ballone-Jones-S} simulations.
Though the $\lambda$-transition in S has been known for over a century, it still attracts a lot of attention and was recently studied by several experimental techniques including
neutron diffraction \cite{Bellissent}, Raman spectroscopy \cite{Kalampounias}, inelastic X-ray diffraction (XRD) and acoustic dynamics \cite{Monaco} or infrared photon correlation spectroscopy \cite{Scopigno}.

For an equally long time, another transformation in liquid sulfur ($l$-S) has been indicated by the change in the color of the heated melt, which turns
from deep-yellow at $T_{\lambda}$, through dark-orange and dark-red to dark-brown-red when approaching the boiling point (445 $\degree$C). By this point, sulfur
melt becomes highly reactive, which was attributed to the presence of various short molecules with free radicals at chain endings responsible for melt color and
reactivity \cite{Meyer, Steudel-Eckert-S-review, Eckert-Steudel-S-review}. This also coincides with the fact that short molecules with length 2-8 are constituents of sulfur vapor,
which has in fact the same color as hot $l$-S in the vicinity of the boiling point \cite{Meyer, Eckert-Steudel-S-review}. Different colors on Jupiter's moon Io - ranging from yellow to
dark-brown are probably also caused by these short molecules on Io's surface \cite{Steudel-Eckert-S-review, Eckert-Steudel-S-review}.
This structural transformation thus can be characterized as a progressive breaking of long chains into short chain-like fragments and the transformation
sequence in $l$-S at ambient pressure then proceeds from S$_8$-molecular to long-polymeric and finally to short-polymeric state. This liquid state evolution
is directly reflected e.g. in viscosity, which has minimum below $T_{\lambda}$ \cite{Steudel-Eckert-S-review} (in molecular liquid), then rises sharply above 159 $\degree$C
(with the onset of polymerization) reaching maximum around 180 $\degree$C \cite{Steudel-Eckert-S-review, Scopigno} (when the average chain length is maximal) and then
again gradually decreases \cite{Steudel-Eckert-S-review} (as short and more mobile molecules emerge). At viscosity/chain-length maximum, other physical properties
like density and electrical conductivity acquire minimal values \cite{Steudel-Eckert-S-review}. Average length of the polymeric units in $l$-S was first directly
measured by electron spin resonance by Gardner and Fraenkel and it was found to be of order $10^4$ at 171$\degree$C \cite{Gardner-Fraenkel} and then decreasing with temperature.
Theoretical models \cite{Tobolsky-Eisenberg, Wheeler-Kennedy-Pfeuty} and computer simulations \cite{Tse-Klug-S, Ballone-Jones-S} also qualitatively reproduce this temperature-induced chain shortening with longest chains near $T_\lambda$.

Recently, Liu \textit{et al.} \cite{Liu-S} reported observation of chain-breakage phenomenon - a rather sharp structural change in heated $l$-S at 6 GPa and 1000-1100 K between forms containing
very long ($\sim$600 at 1000 K) and short ($\sim$19 at 1100 K) chains. We will refer to these two forms of $l$-S as LTL and HTL, respectively.
This experiment was the first direct structural study of heated $l$-S under high pressure, which apparently exhibits an abrupt decrease in average chain length in contrast to
continuous ambient-pressure evolution \cite{Meyer, Steudel-Eckert-S-review, Eckert-Steudel-S-review, Tse-Klug-S, Ballone-Jones-S, Gardner-Fraenkel, Tobolsky-Eisenberg, Wheeler-Kennedy-Pfeuty}.
This process also seems to be related to earlier experiments of Brazhkin \textit{et al.} \cite{Brazhkin-S, Brazhkin-S-Se-I} who reported a nonmetal-metal
transition eventually consistent with chain breakage \cite{Liu-S}. However, in Brazhkin's study, one additional transition in nonmetallic state of $l$-S was indicated.
Very recently, Zhao and Mu \cite{Zhao-Mu} performed \textit{ab initio} MD simulations and reported two successive chain breakages in accordance with all experimental data.

In this paper, we report results of our \textit{ab initio} MD study of $l$-S at 10 GPa that include fragmentation of S$_8$ molecules into short chains on heating and 
polymerization of these fragments into long chains upon subsequent cooling, in accordance with Ref. \cite{Liu-S}. Also, we performed simulations along 1085 K isotherm to
compare our results with recent simulations \cite{Zhao-Mu}. We conclude that there is only one chain-breakage phenomenon in sulfur at investigated conditions - between long/infinite chains
and very short chains of only a few atoms. Our results can be directly associated with the experimentally observed chain-breakage phenomenon \cite{Liu-S} and semiconductor-metal (SC-M)
transition \cite{Brazhkin-S,Brazhkin-S-Se-I}. On the other side, we did not confirm transition in semiconducting state of $l$-S \cite{Brazhkin-S,Brazhkin-S-Se-I}, which was associated
with one additional chain breakage \cite{Zhao-Mu}. We provide a structural description of the chain-breakage process and discuss several structural and physical properties in sulfur HTL
and LTL phases.

The paper is organized as follows. After describing the methods and simulation protocol, we turn to results, which include:

(i) short and medium-range order structural characteristics of chains - coordination number and average lengths, cross-linking and angular and dihedral angle distributions

(ii) physical properties - self-diffusion coefficient, electronic transition and configurational entropy

(iii) comparison of HTL-LTL transformation in $l$-S to similar phenomena occurring in liquid Se and Te

Throughout the text, we compare our results to experiments and simulations already performed, whenever possible. Finally, some open questions
like possibility of another transition in $l$-S and existence of critical point(s), in connection to much older studies, are discussed.

\section{Simulation methods}

All \textit{ab initio} MD simulations were performed with density functional theory (DFT) based VASP 5.3 code \cite{VASP-1, VASP-2}
employing projector augmented-wave pseudopotential (with six valence electrons for S) and Perdew-Burke-Ernzerhof (PBE)
parametrization of the exchange-correlation functional. To simulate \textit{NPT} ensemble, we used VASP 5's implemented Parrinello-Rahman
barostat \cite{PR-2} running with Langevin stochastic thermostat.

Accurate description of the investigated system within DFT would include use of full spin-polarized calculations \cite{Kresse-Se, Shimojo-Hoshino-Zempo}, since
high-temperature liquid S and Se were experimentally found to be paramagnetic at low pressures. This follows from the fact that electron-spin resonance
of $l$-S \cite{Gardner-Fraenkel} and $^{77}$Se nuclear magnetic resonance \cite{Warren-Dupree} were in fact used to track down the concentration of
unpaired electron states - free radicals that are localized at chain endings \cite{Shimojo-Se, Hoshino-Shimojo, Shimojo-Se-2, Shimojo-Hoshino-Zempo, Kresse-Se}
and thus are directly related to average lengths of chain.
Such estimated lengths in $l$-S \cite{Gardner-Fraenkel} and $l$-Se \cite{Warren-Dupree} decrease with temperature as bonds break and paramagnetic susceptibility increases.
However, Warren and Dupree \cite{Warren-Dupree} revealed that this effect is particularly demonstrated only at low pressure, where spin-density centers remain well-separated, like
in $l$-Se near zero pressure or in Se vapor \cite{Warren-Dupree}. Under application of only a few bars, magnetic moments delocalize, changing from strong Curie to weaker Pauli
paramagnetism, and induce (full) metallization \cite{Warren-Dupree}.

This observation in $l$-Se \cite{Warren-Dupree} agrees with our testings in S, where we have found that spin-polarization effects are severely suppressed under pressure, as far as
difference in total energy between spin-unpolarized and paramagnetic short-chain phase (modeled by disordered local moments) was found to be negligible at 10 GPa compared to
(also rather mild) $\sim$10 meV.at$^{-1}$ at 0 GPa. This is also in agreement with testing performed by Kresse \textit{et al.} \cite{Kresse-Se}
who obtained very similar energy differences for $l$-Se at 10 bar and suggested that only a small number of chain endings probably represents paramagnetic centers,
concluding that spin-calculations will not have significant influence on structure and forces in $l$-Se \cite{Kresse-Se}.

In Ref. \cite{Shimojo-Hoshino-Zempo}, Shimojo \textit{et al.} also investigated spin effects on supercritical fluid Se, in which short chains (mostly Se$_2$) are fairly distant from each other.
They found that spin-polarization is pronounced, but at very low densities ($\le$1 g.cm$^3$), far from liquid and even more far from compressed-liquid state conditions.
Based on these results, we performed non spin-polarized calculations gaining factor of two in speed.

\section{Results: Structure and physical properties}

We started MD simulations from S$_8$ molecular structure taking $2\times2\times1$ supercell of S-I containing 512 atoms - a fairly
large system needed to adequately describe disordered liquid state within periodic boundary conditions. Since the chain breakage \cite{Liu-S}
and SC-M transition in S \cite{Brazhkin-S,Brazhkin-S-Se-I} were found to be mostly temperature-driven, we performed simulations
along 10 GPa isobar between 700 and 2000 K to compare our results with experiments and also along 1085 K isotherm at 2-12 GPa to make comparison
with recent MD results \cite{Zhao-Mu}. Total simulation time of our study exceeds 400 ps, which is a run long enough to accumulate statistics that together
with the size of the system maintains reliable results.

\begin{figure}
\begin{tabular}{cc}
\fbox{\includegraphics[width=\columnwidth]{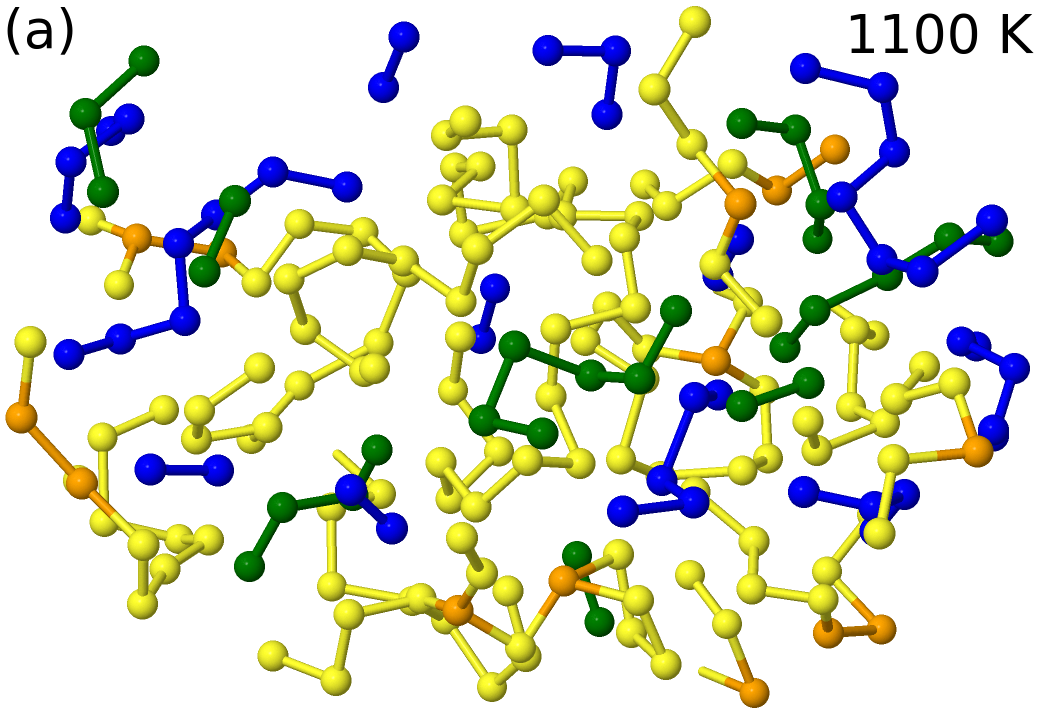}}\\
\fbox{\includegraphics[width=\columnwidth]{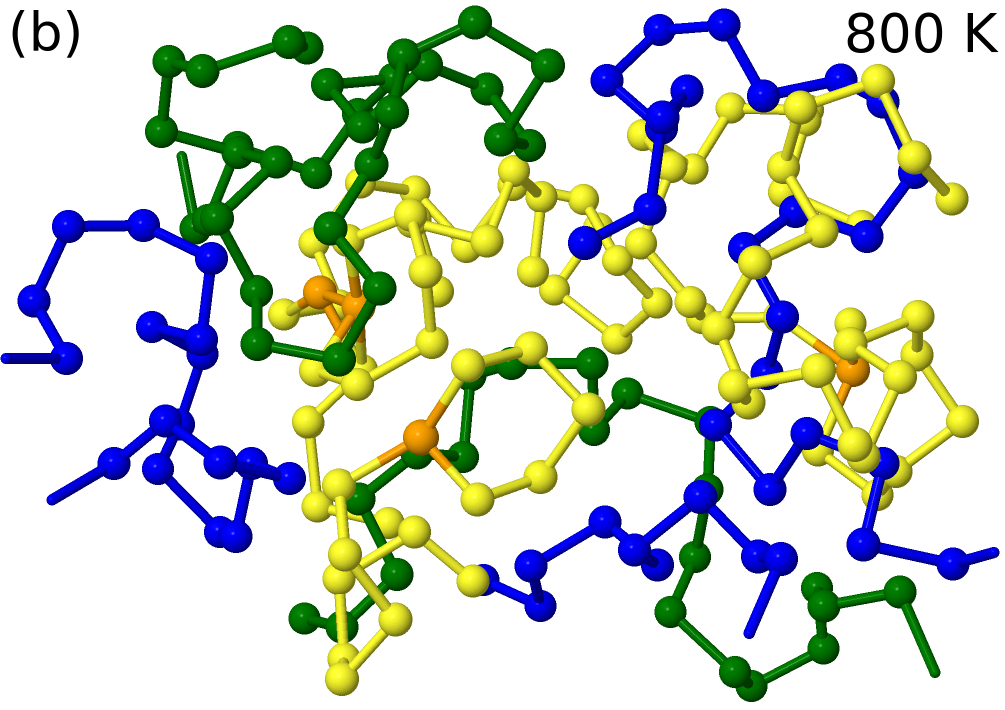}}\\
\end{tabular}
\caption{Liquid sulfur at 10 GPa and 1100 K - HTL form (a) and at 800 K - LTL form (b) - smaller parts extracted from the configuration of the 512-atomic system.
Blue and green chains highlight some individual simple chains and $3c$-S cross-link atoms are marked by orange spheres.
In (a), some typical short chains of HTL are highlighted (others are yellow), while in (b), all long simple chains are marked blue or green.
In both (a) and (b), insides of all branched chains are shown with yellow atoms. The pictures were generated by Jmol \cite{Jmol}.}
\label{HTL-LTL}
\end{figure}

The sample was initially structurally optimized to 10 GPa and then heated to 1000 K in constant pressure-temperature ensemble. After rising the temperature to 2000 K
the system melted and the resulting structure of $l$-S consisted of very short chain fragments. From this point, we gradually lowered the temperature to
1500, 1300, 1200, 1100, 1000, 900, 800 and finally to 700 K \footnote{At 700-800 K, liquid sulfur falls into undercooled region close to melting line.}, where the liquid
contained very long and also infinite (within periodic supercell) chains. We hence observed chain polymerization upon cooling, which represents a reverse reaction to the experimentally
recognized chain breakage induced by heating, and associate findings of our simulations with the reported HTL and LTL forms of $l$-S \cite{Liu-S}.
Snapshot of the simulated HTL phase at 1100 K and LTL phase at 800 K is shown on Fig.~\ref{HTL-LTL} (a) and (b), respectively.

In the simulated system, three types of sulfur atoms were present - single-coordinated ($1c$) representing chain edges, two-coordinated ($2c$)
forming insides of the chains and three-coordinated ($3c$) corresponding to chain cross-links. These atoms together formed two types of chains - simple (isolated),
defined as series of $2c$ sulfur atoms terminated by $1c$-S at both ends, and branched, consisting of $2c$-S inside atoms and ended with one $1c$-S and one $3c$-S atom.
The $3c$-S atoms give rise to branched chains, which were in both HTL and LTL form present along simple chains in a relevant number.
Some closed loops also occasionally emerged, but in negligible proportion compared to chains.

\subsection{Chain lengths and cross-linking}

\begin{figure}
\begin{tabular}{cc}
\includegraphics[width=\columnwidth]{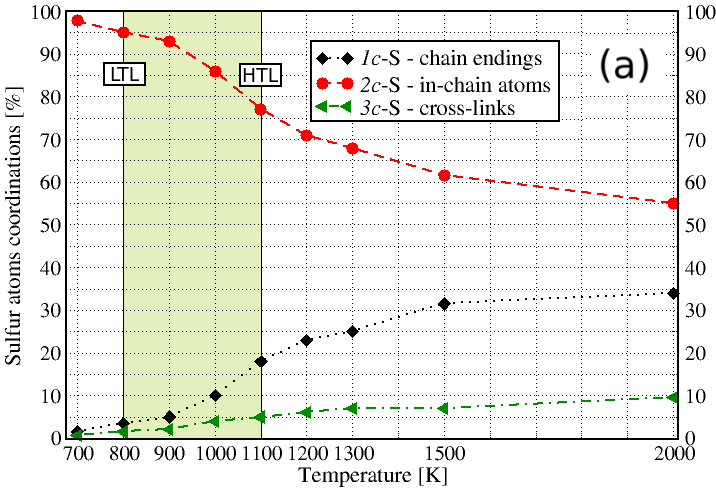}\\
\includegraphics[width=\columnwidth]{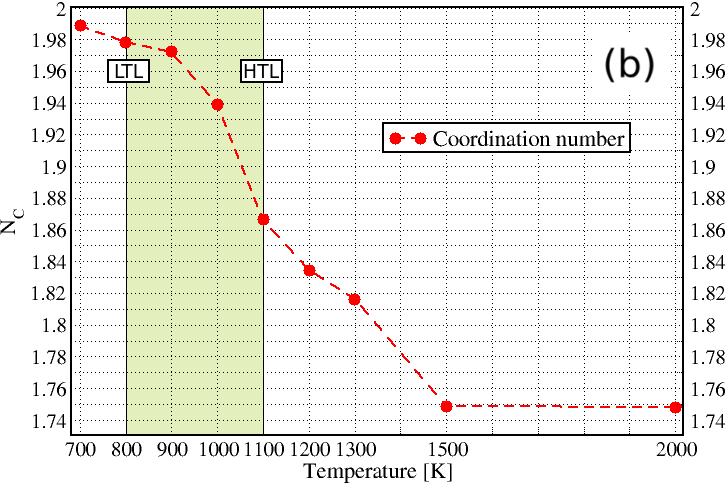}\\
\includegraphics[width=\columnwidth]{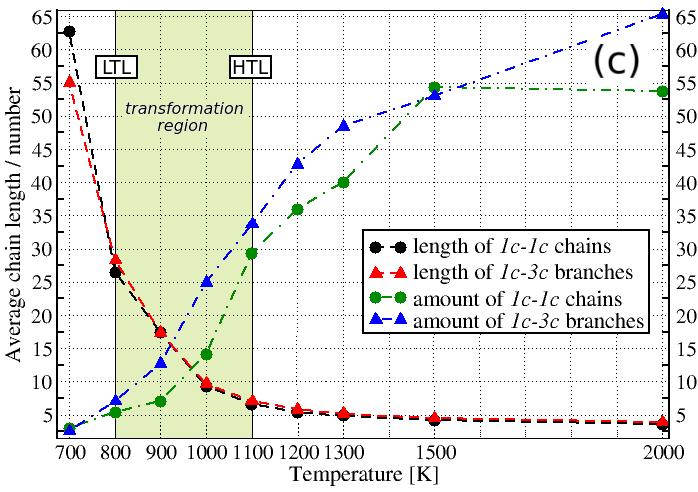}\\
\end{tabular}
\caption{Evolution of S atoms coordinations [\%] (a), coordination number (b) and average length and amount of simple and branched chains (c) during temperature decrease
from 2000 to 700 K at 10 GPa. The LLT between HTL and LTL form proceeds gradually and is most notable between
800 and 1100 K - see the chain length evolution (c). This area is highlighted. Note that the values of average chain lengths at low temperatures
should be even higher due to presence of some infinite chains in the periodic supercell. These were not included in the statistics,
which, however, does not change the qualitative description of the transformation, but suggests that even longer chains are present at low temperatures.
The amount of simple chains and branches remains similar at all temperatures.}
\label{coor-Nc-length}
\end{figure}

The process of polymerization induced by cooling down the reactive HTL sulfur below 1100 K is reflected in temperature dependence of individual
S atoms coordinations depicted on Fig.~\ref{coor-Nc-length} (a). This is closely related to the evolution of coordination number N$_\text{C}$ -
Fig.~\ref{coor-Nc-length} (b) and average lengths of both simple and branched chains - Fig.~\ref{coor-Nc-length} (c). As we can see from the graph, at all temperatures the most common
were $2c$ sulfur atoms (insides) followed by $1c$-S (edges) and some (non-negligible) $3c$-S (cross-links). The abundance of $1c$, $2c$ and $3c$-S shows that
along gradual quenching of the system from 2000 to 700 K, the $2c$-to-$1c$ ratio rapidly increases below 1100 K, which is caused by transformation of short chains
into long polymers as the amount of inside $2c$ atoms increases on behalf of edging $1c$-S. The number of cross-links drops from considerable 9.5\% at 2000 K to a
tiny amount of 0.7\% at 700 K, but since there is only a small total number of chains below 900 K, the relative proportion of simple and branched chains 
remains relatively unchanged and close to 1 (see Fig.~\ref{coor-Nc-length} (c)).

During chain polymerization N$_\text{C}$ increases from approximately 1.75 at 2000 K to 1.99 at 700 K (Fig.~\ref{coor-Nc-length} (b)).
This rather small change in the value of N$_\text{C}$, however, is related to a major structural change in $l$-S, as was proposed by authors of Ref. \cite{Liu-S}.
This change is best described as a transformation between short and long-chain form of polymeric liquid, which is illustrated on Fig.~\ref{coor-Nc-length} (c), where evolution of average chain
length and amount is shown separately for simple $1c$-$1c$ chains and $1c$-$3c$ branches of branched chains ($3c$-$3c$ elements were omitted from the statistics).
Chain lengths were calculated directly from MD trajectories (taking the bond-length limit at minimum of the corresponding pair-correlation function $\sim$2.45 \AA) providing a more direct information than the solely N$_{\text{C}}$-based estimations presented in Refs. \cite{Liu-S,Zhao-Mu}, which do not take into account the branching sites. The phenomenon of chain breakage taking place upon heating, or equally chain polymerization induced by cooling, can be clearly identified from the graph, where average lengths increase from less than 5 in HTL regime to $\sim$60 (lower limit) in LTL.

Existence of branched chains in high-temperature liquid sulfur, which we directly see in our simulations, was already proposed theoretically by recognizing that cross-link elements
(and resulting branches) lead only to a minor energy increase compared to unbranched chains \cite{Wong-Steudel-Steudel}. $3c$ atoms were also observed in previous first-principles MD
simulations of $l$-S \cite{Tse-Klug-S}, $l$-Se \cite{Kirchhoff-1, Kirchhoff-2, Kirchhoff-3, Kresse-Se, Bichara, Hohl-Jones-Se} and also $l$-Te \cite{Akola-Jones-1, Akola-Jones-2}, although in other studies it was observed that $3c$-Se defects are unstable \cite{Hoshino-Shimojo, Shimojo-Se, Shimojo-Se-2}.

In the study of Liu \textit{et al.}, authors measured structure factors, from which they extracted radial distribution functions (RDFs) and
calculated N$_\text{C}$ - Fig.~4 (b) of Ref. \cite{Liu-S}. Thereafter, by taking the simple chain model, which is based on the assumption that all
atoms are either $1c$ or $2c$, they estimated average chain length $L$ from the formula N$_\text{C}$ = 2-2/$L$. Based on the graph of $L$ evolution from their
measurements - Fig.~4 (c) of Ref. \cite{Liu-S}, authors identified liquid-liquid transition occurring between 1000-1100 K at 6 GPa.
Our results shown on Fig.~\ref{coor-Nc-length} (b) and (c) can be directly compared to the experimental data presented on Fig.~4 (b) and (c) of Ref. \cite{Liu-S}, respectively
and even though our values are constrained by the finite simulation size, we see that average chain lengths actually agree very well with lengths estimated from structure factors \cite{Liu-S}.
The only difference is present at higher temperatures, where in the experiment N$_\text{C}$ further progressively decreased and at 1700 K it was proposed that HTL state contains some
amount of monoatomic sulfur (as indicated by N$_\text{C}<1.0$) \cite{Liu-S}, which was not reproduced in our MD study.

\begin{figure}
\includegraphics[width=\columnwidth]{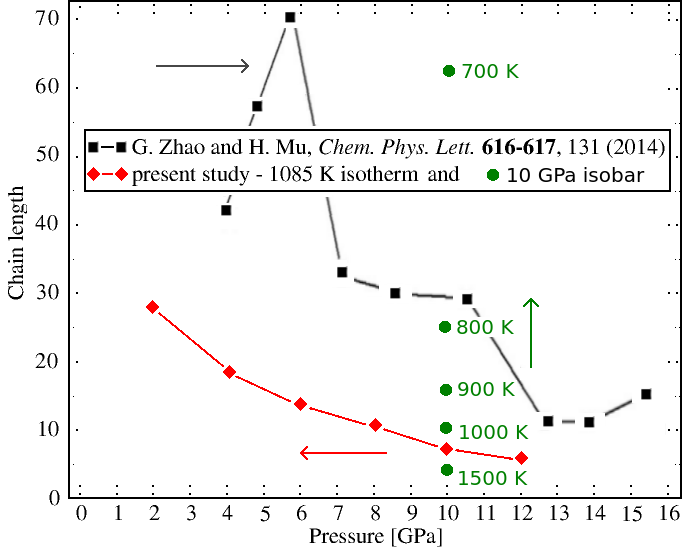}
\caption{Evolution of average length of simple chains obtained from our simulations during decompression from 10 GPa at 1085 K (red diamonds)
and upon cooling at 10 GPa (green circles - the same data as on Fig.~\ref{coor-Nc-length} (c)) from the HTL phase,
along values from Ref. \cite{Zhao-Mu} (black squares).}
\label{lengths}
\end{figure}

In the recent work of Zhao and Mu \cite{Zhao-Mu} the authors claimed that they have observed two successive chain breakages at 1085 K at 6.5 and at 12 GPa, in an apparent accordance with Brazhkin’s experiments \cite{Brazhkin-S, Brazhkin-S-Se-I}. In our study, however, we did not confirm this prediction. In Fig.~\ref{lengths} we plot average chain lengths obtained from our simulations on decompression from 10 GPa at 1085 K and during cooling at 10 GPa from the HTL phase alongside of the data by Zhao and Mu \cite{Zhao-Mu}. We suggest that all our data presented in Fig.~\ref{lengths} are consistent with a single transformation between short-chain HTL and long-chain LTL phase, which in computer simulations proceeds gradually, instead of two sharp changes involving an intermediate form. Possible reasons for finding apparently two transitions in Ref. \cite{Zhao-Mu} may include insufficient equilibration and system size, since distinctly smaller sample and shorter simulation times were used compared to our study.
Moreover, simulating along isotherm might suppress entropy effects responsible for the temperature-induced transition (transformation is not well manifested upon change of pressure) and lead to slower equilibration because system remains in the transition region, where neither short, nor long chains are markedly preferred. Finally, chain lengths were estimated in Ref. \cite{Zhao-Mu} from the same simple chain model as was used in analysis of the experimental data - only from N$_\text{C}$. This may not accurately correspond to true lengths in MD trajectories. Indeed, taking a look at Figs. 3 (b) and 4 of Ref. \cite{Zhao-Mu}, one can estimate that presence of $3c$-S between 7-11 GPa may have led to artificially increased values of chain lengths, because $3c$-S atoms increase the value of N$_\text{C}$. Actual length of chains at these pressures then should be lower, which would correspond to the HTL phase observed experimentally \cite{Liu-S} and in our simulations. This is supported by observation that relative amount of different S coordinations in our simulations at 10 GPa and 1100 K (Fig.~\ref{coor-Nc-length} (a)) agrees perfectly with coordinations at 10 GPa and 1085 K depicted on Fig. 4 of Ref. \cite{Zhao-Mu}, but at the same time, our presented average lengths of sulfur chains at these conditions are $\sim$7, while in Ref. \cite{Zhao-Mu} are as high as $\sim$30.

\subsection{Chain geometry}

In a free sulfur helical chain, repulsion of electron lone pairs leads to minimal energy conformation with S-S-S angle 106$\degree$ and dihedral S-S-S-S angle
85.3$\degree$\cite{Meyer}. Fibrous sulfur, which is after-the-$\lambda$-transition quenched melt (polymeric sulfur glass formed by parallel helices \cite{Donohue}),
yields similar values - angle 109$\degree$ and torsion 86.5$\degree$ \cite{Springborg}. Under rising pressure, bulk molecular sulfur reaches the polymeric region of stability,
where it further transforms from S-II to S-III, which is accompanied by change in the torsion angle from 97.5$\degree$ in S-II at 5.8 GPa to 46.7$\degree$ in S-III at 12 GPa \cite{Deg-1}
(S-S-S angle remains around 104$\degree$). This allows system to pack in a denser structure.

\begin{figure}
\includegraphics[width=\columnwidth]{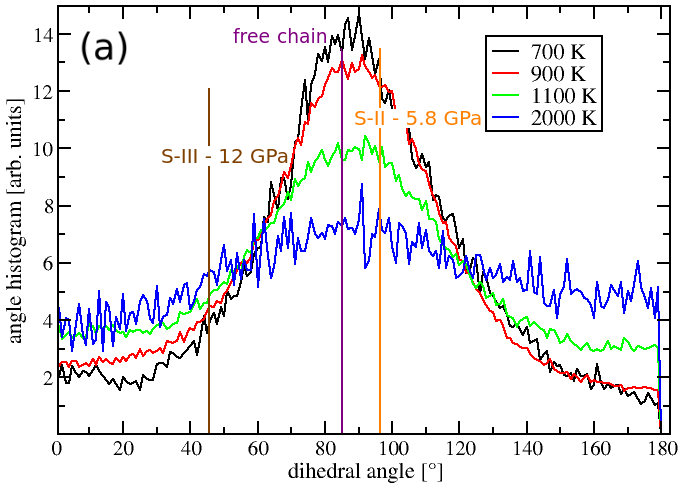}
\includegraphics[width=\columnwidth]{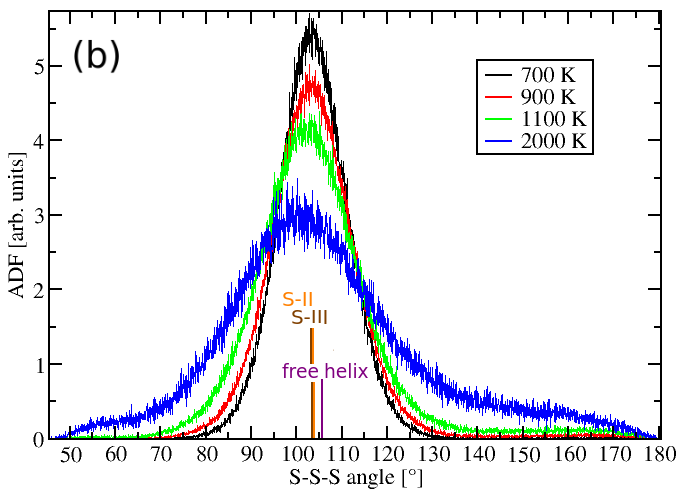}
\caption{Dihedral angle histograms (a) and ADFs (b) of $l$-S at 700, 900, 1100 and 2000 K at 10 GPa. The distributions broaden with temperature considerably, while maxima
shifts are practically negligible. Torsion angles of free helix and S-II and S-III chain phases are marked for comparison.}
\label{dihedrals}
\end{figure}

Regarding our simulated liquid phase, analysis of dihedral angles shows that the distribution in the LTL phase of long chains at 700 K has relatively sharp maximum between
86-90$\degree$ - Fig.~\ref{dihedrals} (a), which was found to be independent on length and type (simple, branched or infinite) of the analyzed chain.
The dihedral conformation in sulfur melt, especially in the long-chain LTL form thus remains very similar to the one found in free helices or chains in S-II.
In all cases, geometry of polymeric units tends to be governed mainly by minimizing energy of electron
configuration. In the HTL form, where the average length of molecules drops below 5-6, the dihedral angle distribution flattens considerably, because the effect of lone-pair repulsion cannot be fully expressed,
as far as in shorter chains there is greater proportion of $1c$-S terminating and $3c$-S branching sites that acquire different geometry.
At the same time, the distribution suffers from poor statistics (scatters), as only a small number of atomic arrangements statistically contributes (there are many chains shorter than 4).
Only absolute values of dihedral angles are shown on Fig.~\ref{dihedrals}, because in liquid state, there is an irregular sign pattern of dihedral angles. This is due to varying orientation of 
chains (see discussions in Refs. \cite{Hohl-Jones-Se, Akola-Jones-1, Akola-Jones-2}), which at high temperatures polymerize in random directions
and do not form regular spiral helices (with +++ or - - - motif) or rings (+ - + - motif) as in crystalline phases.
Analysis of angular distribution functions (ADFs) of bond angles - Fig.~\ref{dihedrals} (b) shows that fairly sharp maximum is placed around 103$\degree$ at 700 K in the LTL form and broader
distribution is characteristic for HTL phase at 2000 K. The location of the maxima still remains close to values typical for all chain forms considered (free, glassy-fibrous and compressed
crystalline S-II and S-III).
We thus found that our simulated liquid phase of sulfur at 10 GPa and especially at lower temperatures, where chains are longer, maintains geometry quite similar to free sulfur helices,
which are rather similar to chains of S-II. This observation suggests that chains in S are deformed only at substantially higher pressures ($\sim$ 40 GPa in S-III)
in order to allow for more efficient packing maintaining higher density.

\subsection{Diffusion}

The nature of the HTL-LTL transformation in $l$-S lies in chain breakage and is therefore likely to have considerable effect on diffusion. To confirm this
assumption, we calculated self-diffusion coefficient $D$ from the Einstein relation
\begin{equation*}
\begin{split}
D = \lim_{t \to \infty}\frac{1}{6t}\Delta r^2(t);\\
\Delta r^2(t) = \left< \left(\bm{r}(t) - \bm{r}_0 \right)^2 \right>
\end{split}
\end{equation*}

and obtained values presented in Table~\ref{tab:D}. A radical change in $D$ takes place after transformation from HTL to LTL, when nearly a two orders of magnitude change takes place
between 1100 and 700 K - at the boundaries of the transformation window.
At 1100 K, our $D$ corresponds perfectly to the value presented by Zhao and Mu at 1085 K and 10 GPa \cite{Zhao-Mu}. These results reflect different movement mechanisms
available for short and long-chain states: long chains are severely constrained and most possibly contain entanglements \cite{Monaco} (connected to rubbery behavior \cite{Monaco, Scopigno}),
while short chains flow much more freely in a diffusion-like manner similarly to molecular liquids. In real systems, possibly both cross-links and entanglements will affect the dynamics of the polymeric liquid.
Analogous arguments may be applied to viscosity, but due to the limited accessible time scale of \textit{ab initio} MD, calculation of viscosity from Einstein relation or Green-Kubo formula - stress
autocorrelation function \cite{Lee-Chang} were out of the reach of the present study.

\begin{center}
\begin{table}[h]
\begin{tabular}{|l|c|c|c|c|c|c|c|c|c|c|}
\hline
$T$ [K], $P$=10 GPa & 700 & 800 & 900 & 1000 & 1100 & 1300 & 1500 & 2000 \\ \hline
$D$ [10$^{-6}$ cm$^2$s$^{-1}$] & 0.44 & 0.75 & 3.05 & 12.7 & 30.6 & 70.9 & 96.7 & 232 \\ \hline
\end{tabular}
\caption{Calculated values of $D$ between 700 and 2000 K.}
\label{tab:D}
\end{table}
\end{center}

\subsection{Electronic transition}

\begin{figure}
\begin{overpic}[width=\columnwidth]{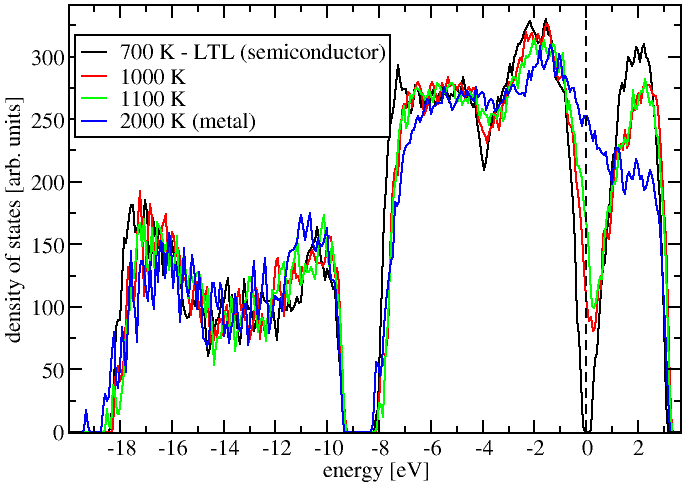}
\put(59,12){\fbox{\includegraphics[width=0.2\columnwidth]{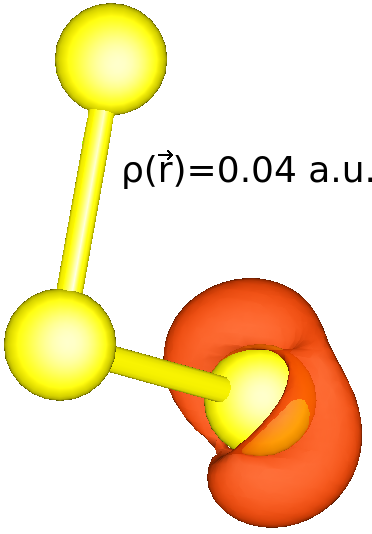}}}
\end{overpic}
\caption{Evolution of electronic DOS at 700, 1000, 1100 and 2000 K throughout the LLT between narrow-gap semiconducting LTL and metallic HTL form at 10 GPa.
Fermi energy is set to zero level, around which progressive filling of the bandgap with rising temperature is well-visible. (Inset) Partial charge density
of electrons with energies in 1.1 eV interval below E$_{\text{F}}$ shown in region near one chain ending - localization at terminating $1c$-S atom is clearly visible.
Density of electrons is plotted as isosurface of value 0.04 a.u.}
\label{DOS}
\end{figure}

Isolated helical chains \footnote{Isolated straight chain of S, which is conducting at ambient conditions, was recently stabilized by constraining it inside carbon
nanotube \cite{Fujimori}.} of sulfur are insulating \cite{Springborg}, while S-II and S-III phases are semiconductors \cite{Nishikawa, Fujihisa}.
Liu \textit{et al.} \cite{Liu-S} suggested that their observed chain breakage in $l$-S might be accompanied by SC-M transition, based on its probable correspondence
to earlier Brazhkin's experiments \cite{Brazhkin-S, Brazhkin-S-Se-I} and from its similarity to the analogous transition present in liquid selenium.
Evolution of density of electronic states (DOS) calculated within the PBE DFT framework (Fig.~\ref{DOS}) shows that at 700 K, the estimated bandgap is
only 0.2 eV (the true one can be larger due to the well-known PBE bandgap underestimation), while above 1000 K, sulfur melt starts to be clearly metallic.
This demonstrates that the short-long chains LLT in sulfur is indeed accompanied by transition between narrow-gap semiconducting LTL and metallic HTL and
we may associate findings of our simulations not only with the observed chain breakage \cite{Liu-S}, but also with the SC-M transition revealed from resistivity
measurements \cite{Brazhkin-S, Brazhkin-S-Se-I}. Semiconducting character of long chains and metallicity of short chains were also confirmed in Ref. \cite{Zhao-Mu}.

Connection between short chains and metallic character of $l$-S is completely analogous to the case of $l$-Se investigated in several studies
\cite{Katayama-Se, Katayama-Tsuji, Tamura, Warren-Dupree, Shimojo-Se, Hoshino-Shimojo, Shimojo-Se-2, Brazhkin-S-Se-I, Brazhkin-Se}. Physical origin of metallization
in $l$-Se was attributed to the presence of chain endings, because electron states with eigenenergies just below the Fermi level E$_{\text{F}}$ - at the top of the valence band, were found to have large
amplitudes around $1c$-Se atoms \cite{Shimojo-Se, Hoshino-Shimojo, Shimojo-Se-2, Shimojo-Hoshino-Zempo, Kresse-Se} (while $3c$-Se were associated with the bottom of the conduction band \cite{Kresse-Se}).
The bandgap then eventually disappears as chains become shorter (and possibly also more branched) with increasing temperature \footnote{This mechanism of metallicity in S and Se contrasts with the SC-M transition in liquid N, where conductivity was found to be strongly correlated with amount of in-chain $2c$-N atoms \cite{Boates-Bonev-N-2009, Boates-Bonev-N-2011}.}. Based on isovalent relationship of S and Se, it is likely that short chain fragments ended by dangling bonds are responsible for metallicity also in HTL form of $l$-S, which we confirmed by plotting partial charge density
in the vicinity of E$_{\text{F}}$ - Fig.~\ref{DOS} (inset).

\subsection{Energy and entropy}

\begin{figure}
\begin{overpic}[width=\columnwidth]{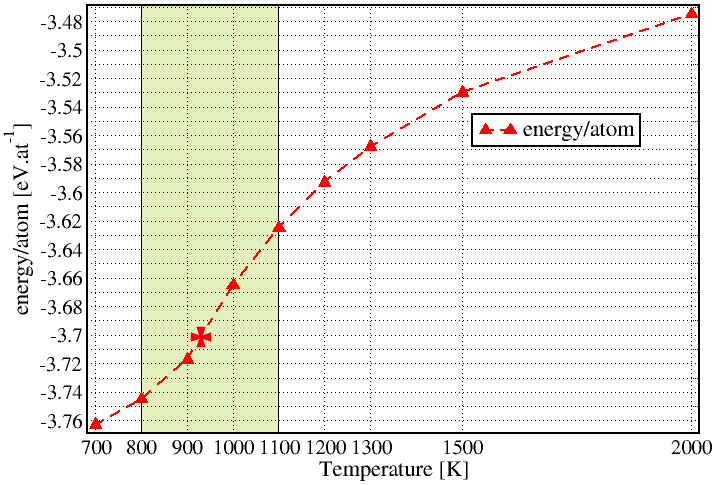}
\put(46,9){\includegraphics[width=0.5\columnwidth]{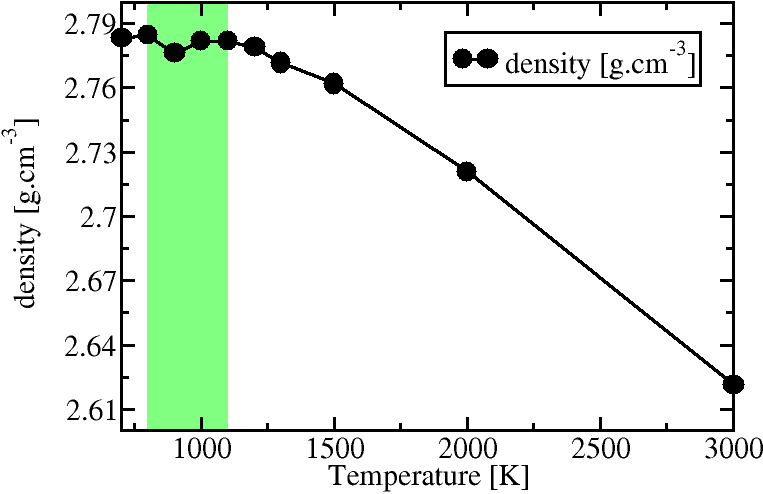}}
\end{overpic}
\caption{Total energy of $l$-S per atom at 10 GPa upon quenching from 2000 to 700 K. An inflection point related to the structural transformation can be recognized near the middle of the highlighted transformation region around 920 K and is marked with cross. (Inset) Density of $l$-S [g.cm$^{-3}$] between 700 and 3000 K.}
\label{energy}
\end{figure}

Since experiments \cite{Liu-S, Brazhkin-S, Brazhkin-S-Se-I} indicate that Clapeyron slope $\frac{dP}{dT}$ of the discussed sulfur LLT is rather low in the $PT$
plane and the transition is dominantly temperature-induced, chain-breakage process must be expected to involve sharp changes in entropy and energy,
though also volume discontinuities were reported for Brazhkin's transitions \cite{Brazhkin-S, Brazhkin-S-Se-I} (but rather small away from triple points).
In our simulations, upon cooling from HTL to LTL state, we observed change in the second derivative of the total energy function, which is apparently related
to the structural change - Fig.~\ref{energy}. At the same time, density remained practically constant 2.78$\pm$0.01 g.cm$^{-3}$ in the
whole transformation region - Fig.~\ref{energy} (inset). From the observed behavior of density and energy, it follows that there must be a decrease in entropy
associated with the HTL-to-LTL process.

From simple considerations, it follows that long chains are preferred in the LTL phase due to their low energy, while higher configurational entropy stabilizes
the HTL form, in which small chain-like fragments have more possibilities how to fill the space. In terms of microscopic theory, this simple description
may be rationalized and numerically estimated from the Flory-Huggins model \cite{Flory}, which provides a crude estimation of mixing entropy in
polymeric-solvent solution. Employing this model for pure polymer (there are no isolated sulfur atoms in our system) yields formula for configurational
entropy $S_c$ per atom
\begin{equation*}
S_c = k \left[ln(Z) + (L-2)ln(Z-1) + (1-L) + ln(L)\right] / L,
\end{equation*}

where $Z$ is coordination of a model lattice (which is fully-filled by disordered chains) and $L$ is some well-defined average length of polymers.
Taking the above formula, the difference in the configurational entropy between the HTL phase of average chain length $\sim$5 and the LTL phase with average length $\sim$60 (see Fig.~\ref{coor-Nc-length} (b)) yields $\Delta S_c = S^{HTL}_c(L=5) - S^{LTL}_c(L=60) \approx$ 0.18 and 0.10 $k$ per atom for typical lattice coordination $Z$ = 6 and 8, respectively. This provides a very rough estimate of the entropy gain per particle in the high-temperature phase, which is due to the increased number of possible configurations. We emphasize again that the Flory-Huggins model corresponds to an incompressible fluid and therefore one cannot expect a quantitative agreement in our case.

The investigated HTL-LTL transformation may be generally viewed as a process, where configurational entropy is reduced as the number of bonds is increased \cite{Corezzi}.
Higher degree of polymerization, which is directly proportional to the average lengths of sulfur chains, maintains lower $S_c$ as far as
within the energy landscape picture \cite{Debenedetti-Stillinger, Stillinger-1, Stillinger-2, Wales}, formation of a new covalent bond (process of HTL-to-LTL polymerization)
may be attributed to the reduction of the number of accessible configurations \cite{Corezzi}. Larger number of bonds in simple systems yields lower entropy.

\subsection{Analogy with selenium and tellurium}

The phase diagram of sulfur shares much analogy with the two other group VI chalcogenes selenium and tellurium. In solid state, the same or similar phases of heavier
Se and Te are stabilized at lower pressures compared to S \cite{Deg-1, Deg-2, Deg-3, Deg-5, Fujihisa, Hejny-McMahon-1, Hejny-McMahon-2}, while liquid
Se and Te exhibit very similar type of structural transformation between short and long(er) chains. Selenium melts directly from polymeric solid into semiconducting
long-chain liquid and then undergoes thermal chain breaking, which also induces metallization. This process follows more or less sharply (depending on
actual conditions) and was studied extensively in several experiments \cite{Warren-Dupree, Katayama-Se, Katayama-Tsuji, Brazhkin-S-Se-I, Brazhkin-Se, Tamura} and simulations
\cite{Kirchhoff-1, Kirchhoff-2, Kirchhoff-3, Kresse-Se, Shimojo-Se, Hoshino-Shimojo, Shimojo-Se-2}. The corresponding solid-liquid-liquid triple point in $l$-Se was estimated
at 3.6 GPa and 900 K \cite{Brazhkin-S-Se-I, Brazhkin-Se}.
Tellurium melts from polymeric solid right to the metallic short-chain form at ambient pressure at around 723 K and upon cooling into
supercooled region, several physical properties exhibit extrema near 626 K \cite{Tsuchiya}. This was also attributed to transformation into form with longer chains and
rings \cite{Akola-Jones-1, Akola-Jones-2}, probably isomorphous to long-chain $l$-Se \cite{Tsuchiya}. These observations show a remarkable similarity in behavior of isovalent elements,
where various solid and liquid phases coincide and their stability regions are progressively shifted to lower-$PT$ conditions with increasing atomic mass.

Another similarity of liquid S and Se, Te lies in geometry of individual chains. In $l$-Se, all studies that included structure characterization reported ADFs with peaks between
103-110$\degree$ \cite{Hohl-Jones-Se, Bichara, Kirchhoff-3, Shimojo-Se}, in accordance with ADFs of our $l$-S - Fig.~\ref{dihedrals} (b). Regarding dihedrals, quite broad distributions
with plateaux in the range from 60$\degree$ to 120$\degree$ (and further flattening on heating) were obtained for $l$-Se \cite{Kirchhoff-3, Hohl-Jones-Se, Shimojo-Se} and are also
very-well comparable to the simulated dihedral distributions of $l$-S in the LTL state, where our distributions are slightly sharper - Fig.~\ref{dihedrals} (a) for 700 and 900 K.
Practically structureless distribution obtained for $l$-Se in the higher-temperature form \cite{Hohl-Jones-Se} again corresponds to the $l$-S HTL phase - Fig.~\ref{dihedrals} (a) for 2000 K.
In amorphous and liquid Te \cite{Akola-Jones-1, Akola-Jones-2}, simulated chains also share similarity with $l$-S in ADFs and (absolute) values of dihedrals, although minor peaks
in dihedral distributions at 0$\degree$ and 180$\degree$ occur in $l$-Te \cite{Akola-Jones-1, Akola-Jones-2}.
However, this effect may probably be attributed to shorter average length of Te chains, which results in greater proportion of chain endings and cross-links with different geometry.

\section{Discussion}

From the results presented so far, it seems clear that the SC-M transformation in liquid sulfur observed back in 1991 by
Brazhkin \textit{et al.} \cite{Brazhkin-S, Brazhkin-S-Se-I} corresponds to the chain breakage phenomenon observed by Liu \textit{et al.},
which we studied as a process of HTL-LTL chain polymerization. However, Brazhkin and colleagues reported two transitions - L-L$^{\prime}$ in nonmetallic
state by means of thermobaric analysis (TBA) and metallization L$^{\prime}$-L$^{\prime \prime}$ (at higher temperatures) indicated by both TBA and
resistivity measurements \cite{Brazhkin-S, Brazhkin-S-Se-I}. While the SC-M transition (which we consider to be associated with L$^{\prime}$=LTL to L$^{\prime \prime}$=HTL transformation)
is well established and could be clearly recognized from anomalies in resistivity, L-L$^{\prime}$ was indicated only by tracking down TBA signals, from which phase
transformation can be identified from a discontinuous change of the (otherwise monotonous) pressure-temperature dependence caused by sudden density change.
Since it was reported that volume changes were rather vague away from the triple points \cite{Brazhkin-S, Brazhkin-S-Se-I} and L-L$^{\prime}$ was not reproduced by
viscosity measurements \cite{Terasaki}, its existence remains unconfirmed \footnote{one could speculate if formation of partially monoatomic $l$-S at 1700 K \cite{Liu-S} could
not possibly correspond to one additional of Brazhkin's transitions, but this would contradict the association of LTL-HTL chain breakage with SC-M transition,
because the short-monoatomic transformation would have to occur inside the metallic region}.
Though Zhao and Mu \cite{Zhao-Mu} associated L-L$^{\prime}$ and L$^{\prime}$-L$^{\prime \prime}$ transitions with two successive chain breakages,
our study does not support this view.

The very interesting observation is that at atmospheric pressure, transformation from long chains into short chain-like molecules in sulfur melt has been already
known experimentally \cite{Meyer, Eckert-Steudel-S-review, Steudel-Eckert-S-review, Gardner-Fraenkel} and
theoretically \cite{Tobolsky-Eisenberg, Wheeler-Kennedy-Pfeuty, Tse-Klug-S, Ballone-Jones-S} for a very long time as a milder process of temperature-induced average chain
length decrease, instead of sudden chain breakage. One can therefore speculate that there might be a critical point somewhere in the lower-pressure region of $l$-S that terminates the
possibly first-order high-pressure LLT, which is manifested in a continuous way at pressures near ambient.
The idea of critical regime is not new and was already presented by Brazhkin \textit{et al.}, who reported that during the SC-M transition in $l$-S,
the observed resistance jump becomes sharper with increasing pressure \cite{Brazhkin-S, Brazhkin-S-Se-I}. On pressure decrease, both TBA and resistivity anomalies
were found to become smaller and finally could not be seen at all below 4-4.5 GPa, where neither L-L$^{\prime}$, nor L$^{\prime}$-L$^{\prime \prime}$ could be recognized.
Under these conditions, critical points can be expected to emerge, terminating the first-order transformations. In the case of $l$-Se, critical point was also predicted based on
the same character of resistivity-drop dependence \cite{Brazhkin-S-Se-I}.

\section{Conclusions}

We performed first-principles molecular dynamics simulations of high pressure-temperature liquid sulfur in order to analyze the structural transformation between long and short-chain form, which was recently reported in experiment by Liu \textit{et al.} \cite{Liu-S} and simulated by Zhao and Mu \cite{Zhao-Mu}. The transition was observed as a breakdown of long polymers into short chain-like fragments upon heating, which we reproduced as short chains polymerization on cooling. The high-temperature form is composed of chains with average length less than $\sim$5, while at low temperatures the liquid contained long sulfur chains with average length over $\sim$60. In both forms, certain amount of cross-links and considerable abundance of branched chains was observed. Chains in the low temperature phase acquire similar geometry as free sulfur helices and chains present in polymeric phase S-II. Existence of an intermediate state between the long and short-chain form of $l$-S \cite{Zhao-Mu}, proposed to be related to Brazhkin's L-L$^{\prime}$ transition \cite{Brazhkin-S, Brazhkin-S-Se-I}, was not confirmed in our study.

From the structural point of view, the HTL-LTL transformation proceeded in our simulations rather gradually, but from the natural limitations of \textit{ab initio} calculations represented by size and time scales, it was not possible to determine whether the observed changes possibly represent a phase transition and eventually determine its order, since this would require performing finite-size scaling. The structural transformation was accompanied by rather sharp changes of several physical properties including self-diffusion coefficient, electronic structure (SC-M transition) and energy, but not density. The process of chain breakage is clearly entropy-driven, where short chains with higher energy are stabilized by increased disorder of their configuration.

Very similar kind of transformation as we have studied in this paper is exhibited by selenium and tellurium. The transformation process between long and short-chain state of $l$-Se is accompanied by increase of metallic character originating in states localized around chain edges and also by sharp change in diffusion coefficient. Similarly, $l$-Te undergoes structurally-induced semimetal-metal transition in undercooled state, which was also associated with change in average length of rings and chains. Individual chains in these liquid systems are also similar to chains in $l$-S as concluded from angular and dihedral angle distributions.

Some points in behavior of liquid S remain unclear and deserve further research, such as character of the proposed L-L$^{\prime}$ transition inside the semiconducting regime \cite{Brazhkin-S, Brazhkin-S-Se-I} or possibility of the existence of a critical point in heated sulfur melt at low pressures. The last point is particularly intriguing since gradual chain breakage at elevated temperature and atmospheric pressure has been known in liquid sulfur since long time.

\begin{acknowledgments}

This work was supported by the Slovak Research and Development Agency under Contract No.~APVV-0558-10 and APVV-0108-11 and by the
project implementation 26220220004 within the Research \& Development Operational Programme funded by the ERDF. Part of the calculations
were performed in the Computing Centre of the Slovak Academy of Sciences using the supercomputing infrastructure acquired in project
ITMS 26230120002 and 26210120002 (Slovak infrastructure for high-performance computing) supported by the Research \& Development
Operational Programme funded by the ERDF.

\end{acknowledgments}


\providecommand*{\mcitethebibliography}{\thebibliography}
\csname @ifundefined\endcsname{endmcitethebibliography}
{\let\endmcitethebibliography\endthebibliography}{}

\end{document}